\documentclass[showpacs,twocolumn,prd,superscriptaddress]{revtex4-1}
\usepackage{amsmath,amssymb}
\usepackage{graphicx,epstopdf}
\usepackage{graphicx}
\usepackage[usenames,dvipsnames]{xcolor}
\usepackage{slashed}
\usepackage[normalem]{ulem}
\usepackage{bm}
\usepackage{bbm}
\usepackage{bbold}
\usepackage[dvipsnames]{xcolor}

\usepackage[printwatermark]{xwatermark}

\usepackage{hyperref}
\hypersetup{colorlinks=true}

\newcommand{\vphi}{\varphi}
\newcommand{\eps}{\varepsilon}

\newcommand{\mbf}[1]{\mathbf{#1}}
\newcommand{\trm}[1]{\textrm{#1}}
\newcommand{\tsf}[1]{\textsf{#1}}

\newcommand{\be}{\begin{equation}}
\newcommand{\ee}{\end{equation}}
\newcommand{\bea}{\begin{eqnarray}}
\newcommand{\eea}{\end{eqnarray}}

\newcommand{\nn}{\nonumber}

\newcommand{\Sfi}{\tsf{S}_{\tiny{\tsf{fi}}}}

\definecolor{aaaa}{rgb}{0.99, 0.4, 0.01}
\definecolor{bbbb}{rgb}{0.5, 0.3, 0.9}

\newcommand{\figref}[1]{Fig. \ref{#1}}
\newcommand{\figrefa}[1]{Fig. \ref{#1}(a)}
\newcommand{\figrefb}[1]{Fig. \ref{#1}(b)}
\newcommand{\figrefc}[1]{Fig. \ref{#1}(c)}

\newcommand{\eqnref}[1]{Eq. (\ref{#1})}
\newcommand{\eqnrefs}[2]{Eqs. (\ref{#1}) and (\ref{#2})}

\newcommand{\sxnref}[1]{Sec. \ref{#1}}

\newcommand{\vkap}{\varkappa}
\newcommand{\cd}{.}

\makeatletter
\def\ps@pprintTitle{%
 \let\@oddhead\@empty
 \let\@evenhead\@empty
 \def\@oddfoot{}%
 \let\@evenfoot\@oddfoot}
\makeatother

\begin{document}

\title{Pulse envelope effect on nonlinear Compton scattering in electron-laser collisions}
\author{B.~King}
\email{b.king@plymouth.ac.uk}
\affiliation{Centre for Mathematical Sciences, University of Plymouth, Plymouth, PL4 8AA, United
Kingdom}


\date{\today}
\begin{abstract}
Nonlinear Compton scattering is calculated for the collision of an electron with a plane wave pulse. A mid infra-red (IR) peak arises in the photon spectrum due to long-range interference associated with the pulse envelope. The case of a flat-top pulse is studied as a toy model for pulse envelope effects and reduced to two final-state momentum integrations; the case of a sine-squared pulse is studied numerically. A perturbative expansion in charge-field coupling reveals that already at intermediate intensities, many orders are required to correctly capture the structure of the mid-IR peak. By regularising the \emph{classical} result, it is shown that the mid-IR peak is due to plane-wave ponderomotive effects on the pulse envelope.
Finally, it is shown that the mid-IR peak can be isolated using energy, angle and polarisation filters.
\end{abstract}
\maketitle
\twocolumngrid

\section{Introduction}
The amplitude for a probe particle to undergo a given process in a laser pulse is proportional to an integral over all spacetime. When the probability is obtained from this by mod-squaring, one captures interference between the process occuring in different parts of the pulse. Here we will consider nonlinear Compton scattering \cite{nikishov64,kibble64}, which is the emission of a photon by a highly relativistic electron in a classical background of arbitrary intensity. We will assume the interaction with a laser pulse is well-approximated by a plane-wave background of finite duration, as is by now standard  \cite{boca09,seipt11,mackenroth11,Krajewska:2012gc,PhysRevD.99.096018}. Then only the integral over the laser pulse's lightfront direction is non-trivial, and interference can be parametrised using the laser pulse phase.

Various approximations have been devised to define a ``probability rate'', which, when integrated over the entire pulse, gives the total probability for the process to occur. Such a rate is useful, as it can be added to Monte-Carlo generators in Particle-In-Cell simulations of bunches of particles colliding with more complicated (e.g. focussed) laser pulses, potentially describing each particle undergoing multiple processes \cite{elkina11,ridgers12,king13a,bulanov13,ridgers14,Gonoskov:2014mda,blackburn14,gelfer15,Vranic:2016myd,efimenko19,Seipt:2020uxv}. The form of the probability rate approximation can be classified by the length scale of interference that it includes.

In a \emph{constant} plane-wave field the interference between emission at different points in the background cancels exactly, such that the total probability can be written as the integration over the laser pulse phase, of a completely \emph{local}, instantaneous probability rate describing point-like emission \cite{ritus85}. Therefore the emission in ``slowly-varying'' laser pulses is often calculated using a \emph{locally constant field approximation} (LCFA), which takes the completely instantaneous constant field rate and integrates it over the form of the non-constant laser pulse field \cite{nikishov64,ritus85}. (A more precise definition of ``slowly-varying'' has been recently investigated in several works \cite{harvey15,DiPiazza:2017raw,Ilderton:2018nws,DiPiazza:2018bfu,King:2019igt}.) 

A more accurate but less versatile approximation of the spectrum can be acquired by perturbing around a \emph{monochromatic} field. The periodic nature of a monochromatic field includes interference between emission in different cycles of equal amplitude. By including this interference exactly and approximating interference due to neighbouring cycles having \emph{different} amplitudes because of the pulse envelope, one can obtain a \emph{locally monochromatic approximation} (LMA) \cite{Heinzl:2020ynb,cain1,Hartin:2018egj}, which evaluates a monochromatic ``rate'' at the local amplitude of the pulse envelope and integrates it over the entire pulse envelope. Such an approach reproduces harmonic structure in the photon spectrum, which is beyond the LCFA, since the harmonic structure is due to the interference between emission in different cycles. However, the LMA misses the full structure due to the pulse envelope. 

The LCFA and LMA include, respectively, interference effects on subwavelength and wavelength scales. Here, we shall study interference effects in the nonlinear Compton process on the scale of the entire pulse length. The reason that interference can occur on spacetime scales longer than the pulse length is that the electron current does not vanish when the field vanishes. To calculate the radiation generated by the electron when it collides with a finite laser pulse, one must regularise the probability and remove the zero-field contribution. This regularisation must be done in the \emph{classical} case \cite{dinu12} as well as the \emph{quantum} case. The regularisation, which includes interference over arbitrarily large scales, leads to a broad mid-IR peak, which is non-perturbative in the charge-field coupling just like the well-known harmonic structure. Whilst this mid-IR peak can be seen in the numerical results of several authors \cite{Ilderton:2018nws,DiPiazza:2018bfu,Blackburn:2018sfn}, to the best of our knowledge, it has yet to be analysed or discussed in detail. This is the focus of the current paper. 

Interference effects in the nonlinear Compton process at the level of the pulse length have been studied in delta-function pulses \cite{Ilderton:2019vot}, in the shape of ultra-short pulses \cite{Seipt:2016rtk} and in double-pulses \cite{Ilderton:2020dhs}. The IR part of the photon spectrum includes radiative loop corrections \cite{YENNIE1961379,PhysRev.140.B516} and can receive significant contributions from much higher orders of expansion in dressed vertices \cite{Lavelle:2005bt,Lavelle:2006vv,Lavelle:2013wx,Edwards:2020npu,Ilderton:2020gno,Ilderton:2020rgk}, which has contributed to discussion of the Ritus-Narozhny conjecture \cite{Fedotov:2016afw,Podszus:2018hnz,Ilderton:2019kqp,Mironov:2020gbi}. The motivation for wanting to better understanding this part of the photon spectrum in nonlinear Compton scattering is in part due to upcoming high-energy particle-laser experiments such as E320 \cite{Meuren:2020nbw} at SLAC and LUXE \cite{luxe19} at DESY which, by virtue of using low-emittance traditionally-accelerated electron beams, will measure the nonlinear Compton process at a higher precision than has so far been possible. Furthermore, QED will be tested in the intermediate intensity regime of $\xi\sim O(1)$ ($\xi$, also known as ``$a_{0}$'' is the classical nonlinearity parameter equal to the work done by the field over a Compton wavelength in units of the background photon energy \cite{ritus85}), which is neither perturbative nor asymptotic in $\xi$. We stress that the mid-IR peak studied in this paper is \emph{beyond} the LCFA and LMA methods being used to model these experiments.

The paper is organised as follows. Sec. II introduces notation, gives the expressions that will be numerically evaluated, recaps regularisation of $1\to 2$ quantum electrodynamical (QED) processes in plane-wave backgrounds and defines the set-up. In Sec. III, the toy model of a flat-top pulse is presented, which gives an intuitive explanation of much of what will follow for the more realistic sine-squared pulse case. In Sec. IV, the lightfront momentum spectrum for nonlinear Compton scattering in a sine-squared pulse is analysed, with particular emphasis on the emergence of the mid-IR peak as $\xi$ is increased above $\xi=1$, and on the failure of a truncated perturbation expansion to reproduce the entire spectrum. In Sec. V, the problem is analysed using nonlinear classical electrodynamics, and the classical motion of the electron identified that corresponds to the mid-IR peak. In Sec. VI, the angular spectrum and polarisation content of the mid-IR peak is analysed with a view to its separation from the rest of the spectrum, and in Sec. VII the paper is concluded. App. A contains some brief information on the classical calculation. Throughout, we will us a system of units in which $\hbar=c=1$.

\section{Method}
Four-momentum conservation in nonlinear Compton scattering can be written as:
\bea
 p + \bar{n}\vkap = q + k, \label{eqn:pcon}
\eea
where $p$ ($q$) is the incoming (outgoing) electron momentum, $k$ is the emitted photon momentum, $\vkap$ is the laser pulse wavevector and $\bar{n}$ is a continuous number which indicates the momentum contributed to the process by the laser pulse background. Specifically, $\bar{n}=k\cd p/\vkap\cd p(1-s)$, where $s=\vkap \cd k/\vkap \cd p$ is the lightfront fraction of the emitted photon. $s$ is an important variable that will be used in the analysis of the results to parametrise the emission spectrum. For fixed emission angle, varying $s$ is equivalent to varying energy, whereas for fixed energy, varying $s$ is equivalent to varying the emission angle.

The scattering matrix is: $\Sfi = (e/2)\int d^{2}x^{\perp} dx^{+} \,\tsf{S}_{r',r;l}$ where $e<0$ is the electron charge and $\tsf{S}_{r',r;l} = \int d\vphi ~\bar{\Psi}_{r',q} \slashed{A}_{l,k}\Psi_{r,p}$ where $\vphi=\vkap\cd x$ is the plane wave phase, $\Psi_{r,p}$ is the Volkov wavefunction for an electron with spin index $r$, and $A_{l,k}$ is the emitted photon wavefunction in polarisation state $l$. Before giving an explicit expression, we note that our calculation will use photon polarisation states that are perpendicular to the laser wavevector (i.e. $A_{l,k}\cd \vkap=0$), and will mostly use the Boca-Florescu regularisation \cite{boca09}. Then we can write:
\bea
\!\!\!\!\tsf{S}_{r',r;l}\!\! &=& \!\!\tilde{c}\int_{\vphi_{i}}^{\vphi_{f}}\!\!d\vphi \,\bar{u}_{r'}\!\!\left[ \Delta \slashed{\epsilon}^{\ast}_{l} + \frac{1}{2\,\vkap\cdot p}\left(\frac{\slashed{a}\slashed{\vkap}\slashed{\epsilon}^{\ast}_{l}}{1-s}+\slashed{\epsilon}_{l}\slashed{a}\slashed{\vkap}\right) \right]\!\!u_{r}\mbox{e}^{if} \nn \\\label{eqn:Sf1}\eea
\bea
f &=&  \frac{1}{\eta(1-s)} \int^{\vphi}_{\vphi_i} \frac{k\cd \Pi(z)}{m^{2}}\,dz \label{eqn:Srprl1}
\eea
where  $u_r$ ($\bar{u}_{r'}$) are the incoming (outgoing) electron bispinors, where $\eta=\vkap \cd p /m^2$ is the (electron) \emph{energy parameter}, $m$ is the electron mass, $\epsilon$ is the photon polarisation, $\Delta = 1-k\cd \Pi/k\cd p$ is the regularisation factor, $\Pi$ is the classical electron kinetic momentum:
\[
 \Pi = p - a + \vkap \left[\frac{p\cd a}{\vkap\cd p}-\frac{a\cd a}{2\vkap\cd p}\right],
\]
(we use the shorthand $a=eA$ with $A$ as the classical plane-wave pulse four-potential), and $\tilde{c}=[2^{3}V^{3}q^{0}k^{0}p^{0}]^{-1/2}$ is the field normalisation factor. We use this form of regularisation, because we will be considering a so-called ``sandwich'' plane-wave pulse which is defined between two definite phases (lightfront co-ordinates) $\vphi_{i}\leq\vphi\leq\vphi_{f}$ and is zero otherwise \footnote{Calculations were also performed where regularisation was carried out at the probability level using the $i\epsilon$-prescription, and no discernible difference was found in the resulting photon spectra.} \cite{Dinu:2013hsd}. Hence the integration in \eqnref{eqn:Srprl1} is over a finite phase interval.

Since the regularisation terms will play a crucial role, we briefly recap how they are derived (see e.g. \cite{boca09,Ilderton:2019bop}) and thereby show how they encode long-range interference on the order of the pulse duration. The scattering amplitude is an integral over all of spacetime, so the non-trivial integration direction along the laser pulse phase, is infinite. Let us write the exponent in \eqnref{eqn:Srprl1} as $f=f_0+f_a$, where $\lim_{a\to 0} f = f_0$ and consider only the pure phase term in the amplitude (the other parts do not require regularisation). It is useful to split this integral into three pieces around where the pulse has support, which we define to be for $\vphi_{i}<\vphi<\vphi_{f}$:
\bea
 \int_{-\infty}^{\vphi_{i}} \mbox{e}^{if_{0}'x} dx + \int_{\vphi_{i}}^{\vphi_{f}} \mbox{e}^{if(x)} dx + \int_{\vphi_{f}}^{\infty} \mbox{e}^{if_{0}'x+iF} dx,\nn
\eea
where:
\[
 F = \int_{\vphi_{i}}^{\vphi_{f}}f_{a}'(z)dz = f_{a}(\vphi_{f})-f_{a}(\vphi_{i})
\]
is a phase accumulated by the electron from passing through the laser pulse (see e.g. \cite{Ilderton:2020dhs} for how this accumulated phase can impact nonlinear Compton scattering). When evaluated, the two integrals where the pulse has zero support ($\vphi<\vphi_{i}$ and $\vphi>\vphi_{f}$) are assumed to converge as the lower (upper) bound tends to negative (positive) infinity (e.g. through use of the standard ``$i\epsilon$'' prescription \cite{Peskin:1995ev}). Then using integration by parts, the three integrals can be written:
\[
 - \int_{\vphi_{i}}^{\vphi_{f}}\mbox{e}^{if(x)}~\frac{d}{dx}\frac{1}{if'} dx.
\]
Undoing the integration by parts step and rewriting factors, then gives the final result
\[
 \int_{\vphi_{i}}^{\vphi_{f}}\Delta\,\mbox{e}^{if(x)}dx; \quad\Delta = 1- \frac{f'}{f_{0}'}.
\]
Therefore, the $\Delta$ regularisation terms encode the contribution over length scales that are of the order of the pulse duration and beyond.

Using \eqnref{eqn:Sf1}, the total unpolarised probability can then be written as $\tsf{P} = \alpha\mathcal{I}/\eta$, where $\alpha\approx1/137$ is the fine-structure constant and 
\bea
\mathcal{I} &=& \frac{1}{2}\sum_{r'=1}^{2} \sum_{r=1}^{2} \sum_{l=1}^{2} \mathcal{I}_{r',r;l}  \label{eqn:I1}\\
\mathcal{I}_{r',r;l} &=& \frac{1}{2^{4}\pi^{2}}\int_{0}^{1} \frac{ds}{s(1-s)} \int \frac{d^{2}k^{\perp}}{m^{2}} \left|\tsf{S}_{r',r;l}\right|^{2}\nn.
\eea
Later, we will have reason to evaluate specific polarisation channels, so we will not always perform all the polarisation sums in \eqnref{eqn:I1}.
\newline

The example plane-wave pulse background that we will consider, is that of a circularly-polarised sine-squared pulse, which is only defined on a section of the phase line:
\bea 
a=
\begin{cases}
 m \xi \sin^{2}\left(\frac{\vphi}{2N}\right)\left[\eps_{1} \cos \vphi + \eps_{2} \sin\vphi\right] 
 & \trm{if }  \vphi \in [0, 2\pi N]\nn \\
  0 & \trm{otherwise}.\label{eqn:a1}
\end{cases}\\
\eea
$N$ denotes the number of cycles and we choose lab co-ordinates to coincide with $\eps_{1}=(0,1,0,0)$, $\eps_{2}=(0,0,1,0)$ and $\vkap=\vkap^{0}(1,0,0,1)$. Although the amplitude of the pulse and hence its intensity is actually co-ordinate-dependent, we will refer to $\xi$ as the \emph{intensity parameter} in analogy with the monochromatic result. (The monochromatic limit can be reached by letting $N\to \infty$, and extending the pulse to be defined over the entire real line, leading to $-a \cd a \to m^{2}\xi^{2}$.) Our results for the total probability will be characterised by the parameters: $\eta$, $\xi$, and $N$.
\newline

Before studying the effect of the smooth pulse envelope \eqnref{eqn:a1}, we consider a simpler example, for which more analytical progress can be made.

\section{Toy model: flat top pulse} \label{sec:FT1}
Let us define the ``flat top'' pulse by the vector potential:
\bea
a=
\begin{cases}
 m \xi \left[\eps_{1} \cos \vphi + \eps_{2} \sin\vphi\right]
& \trm{if } \vphi \in [0, \Phi]\nn \\
  0 & \trm{otherwise}.
\end{cases}
\eea
Performing the sum in \eqnref{eqn:I1} to obtain the probability for unpolarised particles, the integral can be written \cite{Heinzl:2020ynb}:
\bea
 \mathcal{I} &=& \frac{1}{(4\pi)^{2}\eta}\int d\vphi d\vphi' \frac{ds \,d^{2}\mbf{r}^{\perp}}{m^{2}s(1-s)} \mbox{e}^{\frac{is}{2\eta(1-s)}\int_{\vphi'}^{\vphi} 1 + \frac{(\mbf{r}^{\perp}+s\mbf{a}^{\perp})^{2}}{m^{2}s^{2}}} \nn \\
 &&\quad\times\left[-2m^{2}\Delta \Delta' + g(s)(a^{2}\Delta' + a'^{2}\Delta - 2a\cdot a')\right],\nn\\
 \label{eqn:IFT2}
\eea
where $\mbf{r}^{\perp}=(\mbf{k}^{\perp} - s\,\mbf{p}^{\perp})/m^{2}$ and $g(s)=(1-s + (1-s)^{-1})/2$ and in \eqnref{eqn:IFT2} only, we use the shorthand $a=a(\vphi)$ and $a'=a(\vphi')$. By applying the Jacobi-Anger expansion \cite{watson22} and following a derivation similar to the infinite monochromatic plane-wave case \cite{nikishov64,kibble64,landau4}, both phase integrals and the azimuthal photon momentum integral can be performed analytically, and one arrives at $\mathcal{I} = \sum_{n= n_{\ast}}^{\infty} \mathcal{I}_{n}$ where
\bea
\mathcal{I}_{n} &=& -\frac{\Phi}{2\eta}\int \frac{ds\, d(r^{2})}{s(1-s)} \, \delta_{\Phi}\left[\frac{r^{2}-r_{\infty}^{2}}{2\eta s(1-s)}\right] \nn \\
&& \times\left\{ w^{2}J_{n}^{2}(z) + \frac{\xi^{2}g}{2}\left[2wJ_{n}^{2}(z)-J_{n+1}^{2}(z)-J_{n-1}^{2}(z)\right]\right\},\nn \\ \label{eqn:Inflat}
\eea
with $r = |\mbf{r}^{\perp}|$,
where $z=\xi r / \eta(1-s)$ and we have defined the nascent delta function
\bea
\delta_{\Phi}(x)=\frac{\Phi}{2\pi}~\trm{sinc}^{2}\left(\frac{\Phi x}{2}\right),
\eea
($\trm{sinc}\,x = (\sin x)/x$) and the finite-duration factor $w$ is:
\[
w = \frac{s^{2}+r_{\infty}^{2}}{s^{2}+r^{2}};\qquad r_{\infty}^{2}=2ns\eta (1-s) - s^{2}(1+\xi^{2}).
\]
For a finite pulse, the threshold harmonic $n_{\ast}$ is not constrained, i.e. $n_{\ast} = -\infty$. Compare this to the monochromatic, infinitely-extended result familiar from the literature \cite{nikishov64}, which can be acquired by taking the limit $\Phi \to \infty$ of \eqnref{eqn:Inflat}. In this limit, the background can only contribute momenta in positive integer multiples of the wavevector. For the infinite monochromatic case, the threshold harmonic is:
\[
 n_{\ast} = \lceil \tilde{n}_{\ast} \rceil;\qquad \tilde{n}_{\ast} = \frac{s(1+\xi^{2})}{2\eta(1-s)},
\]
where $\lceil \cdot \rceil$ is the ceiling function. In particular, we note $n_{\ast} \geq 1$ for the infinite monochromatic case. For the flat-top pulse, the $\trm{sinc}$ function represents the spectrum of the finite-duration pulse, which, unlike the infinite monochromatic case, has a finite bandwidth. Instead of the magnitude of the photon's perpendicular momentum, $r^{2}$, being constrained to be a single value, $r_{\infty}^{2}$, it can take a range of values around this, approximately in $r^{2}_{\infty} \pm W$ with $W=\pi [\Phi/2\eta s (1-s)]^{-1}$. (As the $\trm{sinc}$ function has a zero at $\pi$, the bandwidth is taken to be the half-width, at approximately $\pi/2$.) This has a consequence for which harmonics contribute to a given part of the spectrum. This can be seen by writing the finite-width transverse momentum function as:
\[
 \delta_{\Phi}( n- \tilde{n}_{\ast}),
\]
from which we see the harmonic number $n$ has support in the range approximately $n \pm \pi/\Phi$. The most significant difference to the monochromatic case occurs in two parts of the spectrum: i) when close to the end of a harmonic range; ii) at very low values of the lightfront fraction parameter, $s$. Close to the end of a harmonic range, because the infinite monochromatic case constrains the threshold harmonic $n_{\ast}$ to integer values, the spectrum changes abruptly (e.g. the structure of the first harmonic around the highest $s$-value is often referred to as the ``Compton Edge'' \cite{harvey09}); in contrast, for the finite-duration flat-top pulse case, there is no such constraint on $n_{\ast}$  and the spectrum changes smoothly through harmonic ranges. In the infinite monochromatic case, as $s\to 0$, $n_{\ast}\to 1$ because the process  is kinematically forbidden for lower values of $n$, whereas in the finite-duration case, when $s$ is sufficiently small that $\tilde{n}_{\ast} \lesssim 2/\Phi$, the $n=0$ harmonic becomes accessible. In fact, the $n=0$ harmonic also corresponds to a peak, but it originates from the finite bandwidth of the pulse envelope, and not from the carrier phase and is in the mid-IR range of the spectrum. In the same way that the harmonic structure arises from the non-perturbative charge-field coupling, so too does the zeroth harmonic, and its nature will be investigated in the more realistic case of a smooth pulse, in the following sections of the paper. 

\begin{figure}[h!!]
\centering
 \includegraphics[width=0.7\linewidth]{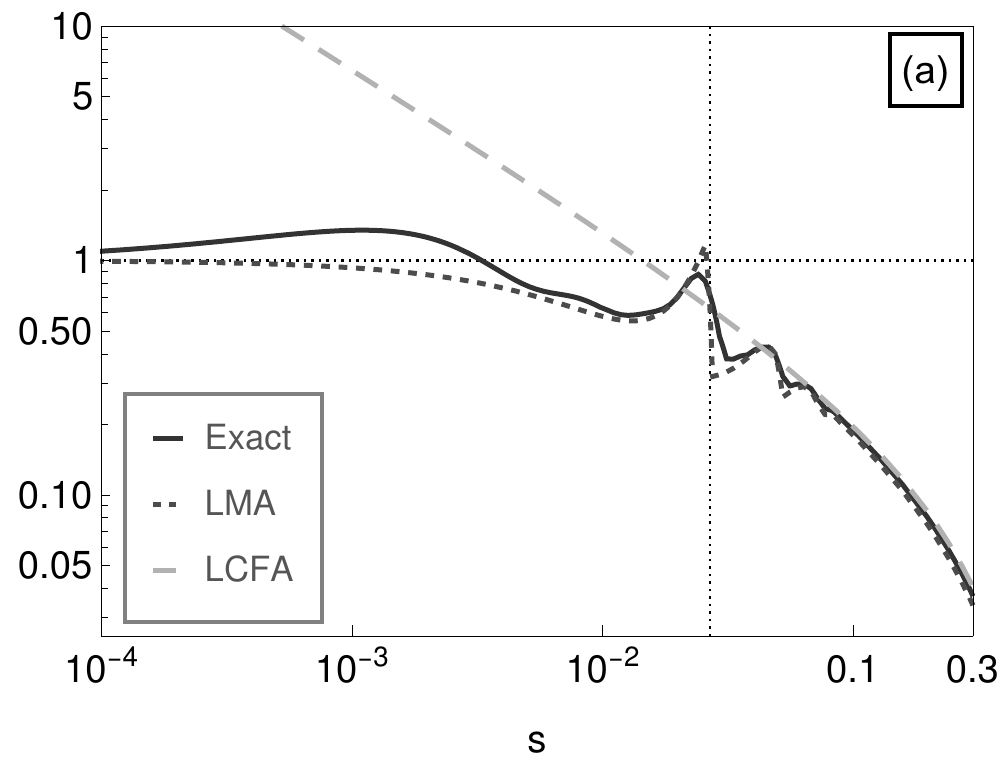}\\
 \includegraphics[width=0.7\linewidth]{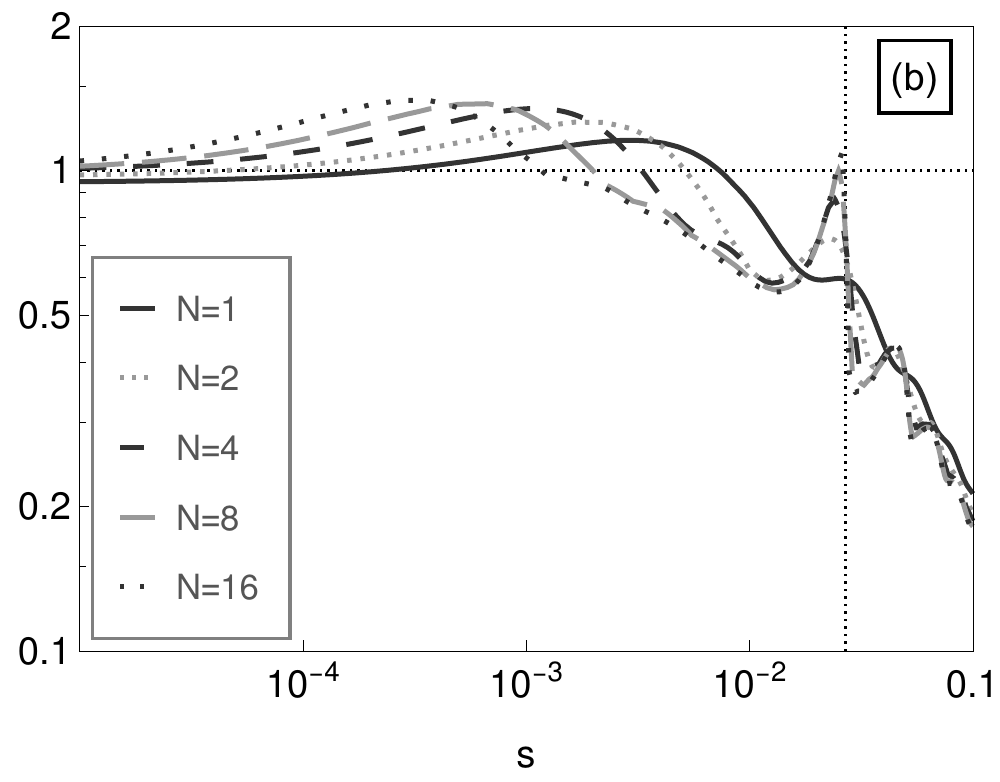}\\
 \includegraphics[width=0.75\linewidth]{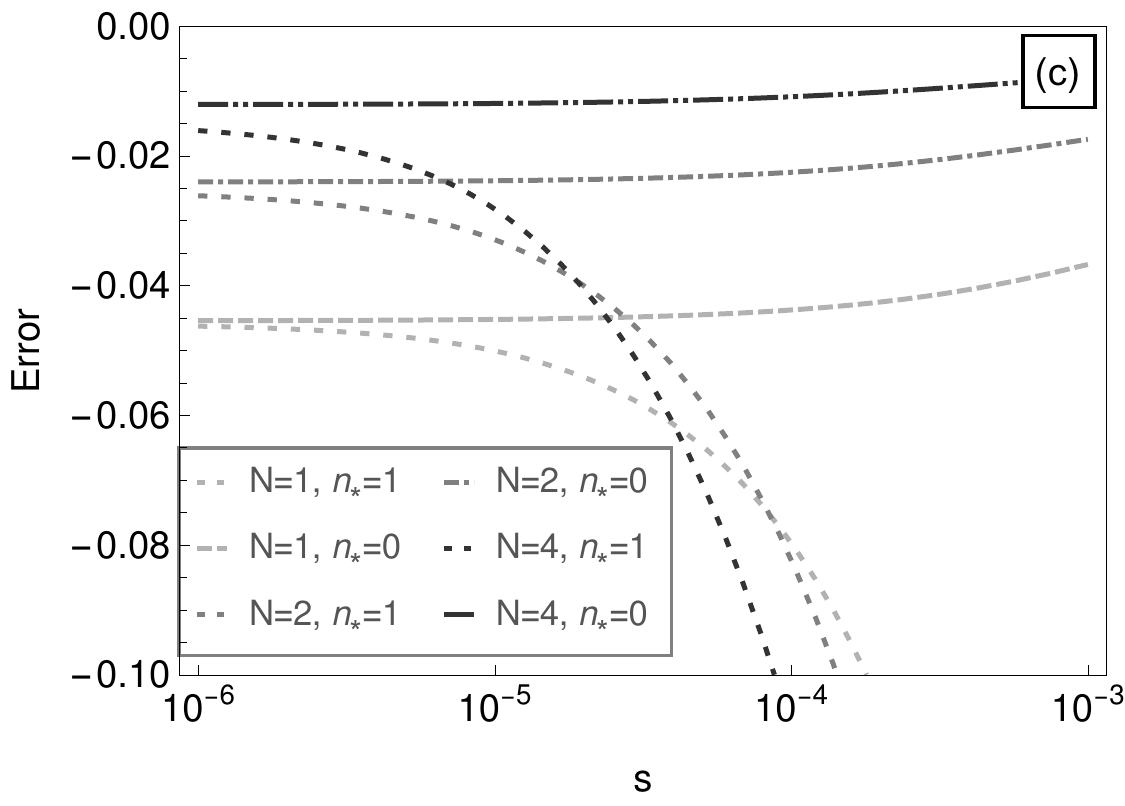}~~~~~
 \caption{Figs. (a)-(b): Compton lightfront momentum spectra ($d\mathcal{I}/ds$) for the flat-top pulse. The amplitude has been normalised to unit IR limit ($\xi^{2}\Phi/2$). $N$ is the number of laser cycles. In Fig. (c), the error in the normalised spectrum is plotted, for when a different lowest harmonic, $n_{\ast}$, is chosen for the harmonic sum: $\mathcal{I}(n_{\ast}=-1)/\mathcal{I}(n_{\ast}) - 1$.} \label{fig:FT1}
\end{figure}

We illustrate the photon spectrum in a flat-top pulse by evaluating \eqnref{eqn:Inflat} and normalising the spectrum to unit IR limit. The spectrum for an intensity $\xi=2.5$, lightfront momentum $\eta=0.1$ and various pulse durations $\Phi=2\pi N$ (where $N$ is the number of cycles) is plotted in \figrefa{fig:FT1} and \figrefb{fig:FT1}. A comparison with the locally monochromatic approximation (LMA) and the locally constant field approximation (LCFA) is made in \figrefa{fig:FT1}, and in \figrefb{fig:FT1}. Two effects of shortening the pulse are demonstrated: the smoothening of harmonic edges and the shifting of the mid-IR peak to lower values of momenta.

We highlight a peculiar property of the flat-top solution: if the pulse is sufficiently short, the $n=-1$ harmonic must also be included, otherwise the IR limit is incorrect. In \figrefc{fig:FT1} we show the error in the spectrum when the harmonic sum is chosen to begin at $n_{\ast}=1$ or $n_{\ast}=0$. This error increases for shorter pulses. Of course, it is kinematically forbidden for an electron to emit two photons (one from the Compton process, and one ``into'' the field in the $n_{\ast}=-1$ harmonic). Instead, it is the case that, due to the finite bandwidth of photons being absorbed from the flat-top pulse, also the $n=-1$ harmonic includes an absorption contribution. 

To conclude this section, we state the classical result for radiation from an electron in a flat-top pulse. $\mathcal{I}^{\scriptsize\tsf{cl.}} = \sum_{n= n_{\ast}}^{\infty} \mathcal{I}_{n}^{\scriptsize\tsf{cl.}}$ where
\bea
\mathcal{I}^{\scriptsize\tsf{cl.}}_{n} &=& -\frac{\Phi}{2\eta}\int \frac{ds\, d(r^{2})}{s} \, \delta_{\Phi}\left[\frac{r^{2}-(r^{\scriptsize\tsf{cl.}}_{\infty})^{2}}{2\eta s}\right] \times\left\{ (w^{\scriptsize\tsf{cl.}})^{2}J_{n}^{2}(z^{\scriptsize\tsf{cl.}}) \right. \nn \\
&&\left.  + \frac{\xi^{2}}{2}\left[2w^{\scriptsize\tsf{cl.}}J_{n}^{2}(z^{\scriptsize\tsf{cl.}})-J_{n+1}^{2}(z^{\scriptsize\tsf{cl.}})-J_{n-1}^{2}(z^{\scriptsize\tsf{cl.}})\right]\right\}, \label{eqn:InflatCl}
\eea
where $z^{\scriptsize\tsf{cl.}}=\xi r / \eta$, where the classical width factor $w^{\scriptsize\tsf{cl.}}$ is:
\[
w^{\scriptsize\tsf{cl.}} = \frac{s^{2}+(r^{\scriptsize\tsf{cl.}}_{\infty})^{2}}{s^{2}+r^{2}};\qquad (r^{\scriptsize\tsf{cl.}}_{\infty})^{2}=2ns\eta  - s^{2}(1+\xi^{2}).
\]
As one might expect, \eqnref{eqn:InflatCl} produces a mid-IR peak that is indistinguishable from the quantum theory, but with harmonic edges at $s^{\scriptsize\tsf{cl.}}_{n} = 2n\eta/(1+\xi^{2})$, which correspond to vanishing electron recoil. The agreement at small $s$ with the quantum theory motivates the classical analysis in \sxnref{sec:CA}.

\section{Lightfront spectrum}
\begin{figure}[h!!]
\centering
 \includegraphics[width=0.9 \linewidth]{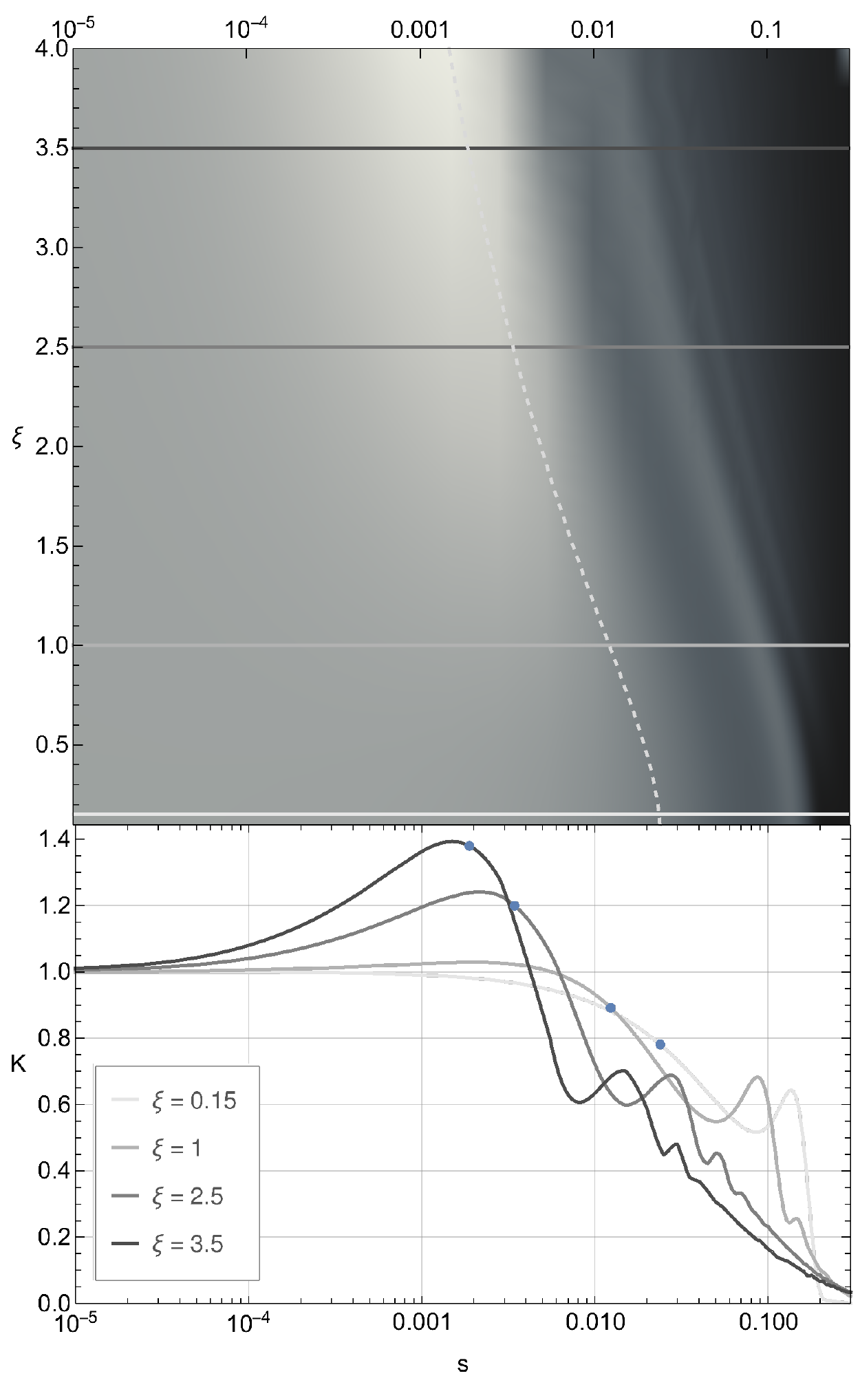}
 \caption{How the mid-IR peak emerges as $\xi$ is increased. The horizontal axis on both figures is identical. The curves in the lower plot correspond to the lineouts indicated by solid lines on the upper plot. The dashed line is the approximate position of the mid-IR peak, which is shown on the lineouts by a dot on each curve. The results are for an $\eta=0.1$ electron colliding head-on with a $4$-cycle ($N=4$) pulse.} \label{fig:midIRpeak}
\end{figure}

We turn now to the more realistic sine-squared pulse envelope from \eqnref{eqn:a1} and illustrate the photon spectrum of nonlinear Compton scattering by evaluating \eqnref{eqn:I1} numerically. We calculate the quantity $\tsf{K}=(1/C)d\mathcal{I}/ds$ for various intensities, where the normalisation factor $C$ is chosen so that $\lim{\tsf{K}}_{s\to 0}=1$, ($C=\frac{1}{2}\int\mbf{a}^{2}d\vphi$, which for the sine-squared pulse is equal to $3N\pi\xi^{2}/8$ \cite{DiPiazza:2018bfu,Heinzl:2020ynb}). This allows for an easier comparison of spectral features. We take $\eta=0.1$, which, for example, for electrons colliding with a $1.55\,\trm{eV}$ (800\,$\trm{nm}$) laser pulse at an angle of $20$ degrees to head-on, corresponds to an electron energy of $8.7\,\trm{GeV}$. The spectrum is plotted in \figref{fig:midIRpeak} for a range of $\xi$ values: $0.1 \leq \xi \leq 4$.
 For small values of $\xi \sim O(0.1)$, the only signifcant structure is the first harmonic and the global maximum is the IR limit. As $\xi$ increases, so too does the relative amplitude of higher harmonics. As $\xi$ is increased above $\xi \approx 1$, we see a new peak develop, which, being  higher than the IR limit, becomes the global maximum. This structure is what we will refer to as the ``mid-IR peak''.  Its appearance only when $\xi \gtrsim 1$ implies that a calculation that is accurate to all orders in $\xi$ is necessary. The position of the mid-IR peak can be approximated by taking the formula for the $n$th harmonic of the monochromatic case, $s_{n}$ where:
\bea
 s_{n} = \frac{2n\eta}{2n\eta+1+\xi^{2}},\label{eqn:sn1}
\eea
setting $n=1$ and rescaling $\eta \to \tilde{\eta}=\eta/2N$ corresponding to a wavevector with wavelength equal to the duration of the envelope. (This value of the approximate bandwidth is equivalent to that in the flat-top pulse toy model studied in \sxnref{sec:FT1}.)  This corresponds to replacing the carrier frequency wavevector, with the wavevector of the envelope (which has a wavelength $4\pi N$, which is $2N$ times larger than for the carrier wave). The position of this approximation is indicated by the dashed line  (dots) in the upper (lower) plot of \figref{fig:midIRpeak}. The mid-IR peak is rather broad, and this method for locating the position does not exactly coincide with the maximum. This is perhaps unsurprising: the flat-top pulse case showed how a broad bandwidth of frequencies arises from the pulse envelope.

\begin{figure}[h!!]
\centering
 \includegraphics[width=0.99 \linewidth]{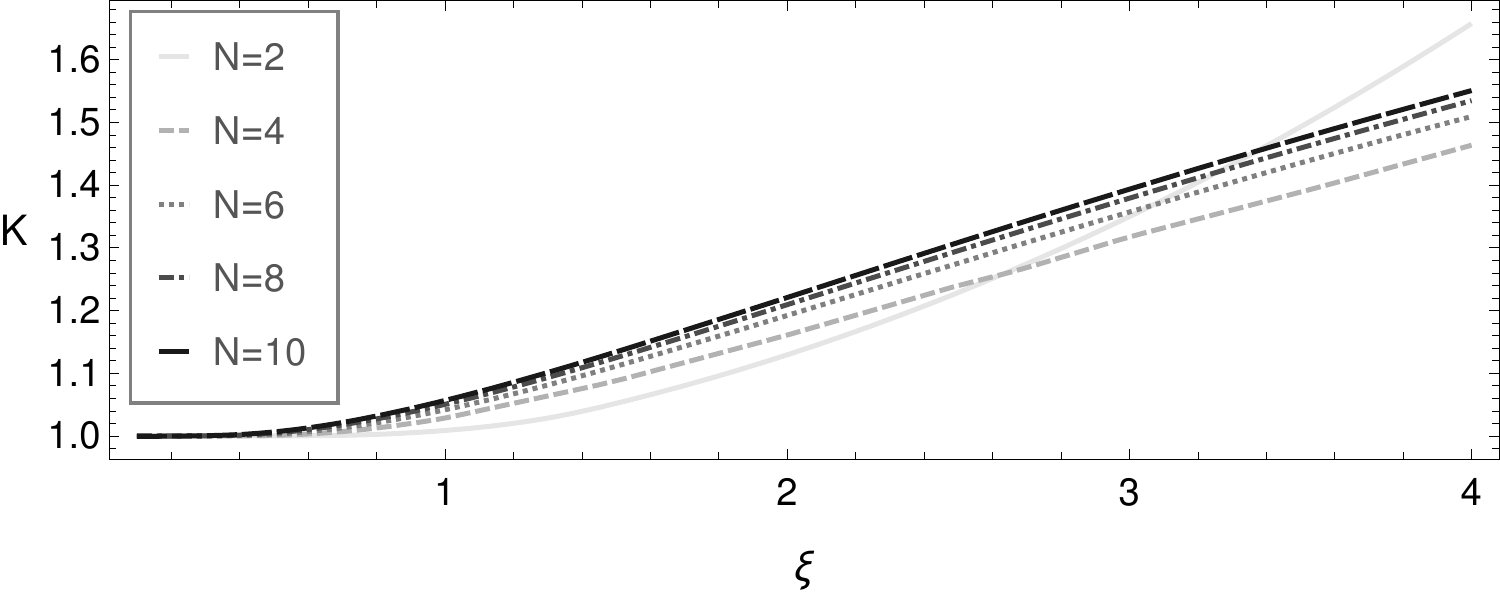}
 \caption{The relative height of the mid-IR peak compared to the IR limit ($\propto \xi^{2}$) grows linearly ($\propto \xi$) when $\xi \gtrsim 1$.} \label{fig:midIRbumpScaling}
\end{figure}

As $\xi$ increases above $\xi \approx 1$, the first harmonic grows as $d\tsf{P}/ds\sim \xi^2$, but the mid-IR peak increases as $d\tsf{P}/ds\sim \xi^3$. This is shown for various pulse durations in \figref{fig:midIRbumpScaling} (numerical results indicate the growth with $\sim\xi^{3}$ is independent of $\eta$). (For $N=2$, the height of the peak appears to grow slightly differently with $\xi$ than it does for longer pulses.) 

To investigate the peak's nonlinear nature, we perturbatively expand $\tsf{K}$ in the nonlinearity parameter $\xi$. Now, it is clear that although the magnitude of the exponential is bounded, the magnitude of a series in $\xi$ of this exponential that is truncated at any finite order, will be unbounded. If the expansion in $\xi$ is performed at the probability level and the $i\epsilon$ method of regularisation is used \cite{Peskin:1995ev}, this leads to difficulties in the evaluation of the zero-field limit pure phase terms. Therefore, we perform the perturbative expansion at the \emph{amplitude level} given in \eqnref{eqn:Sf1}, where the $j$th perturbative order can be understood as including from the background $j$ interactions (``photons'') with the electron. (At the probability level, this only corresponds to consistently including up to order $O(\xi^{(j+1)})$ terms.)

\begin{figure}[h!!]
\centering
 \includegraphics[width=0.9\linewidth]{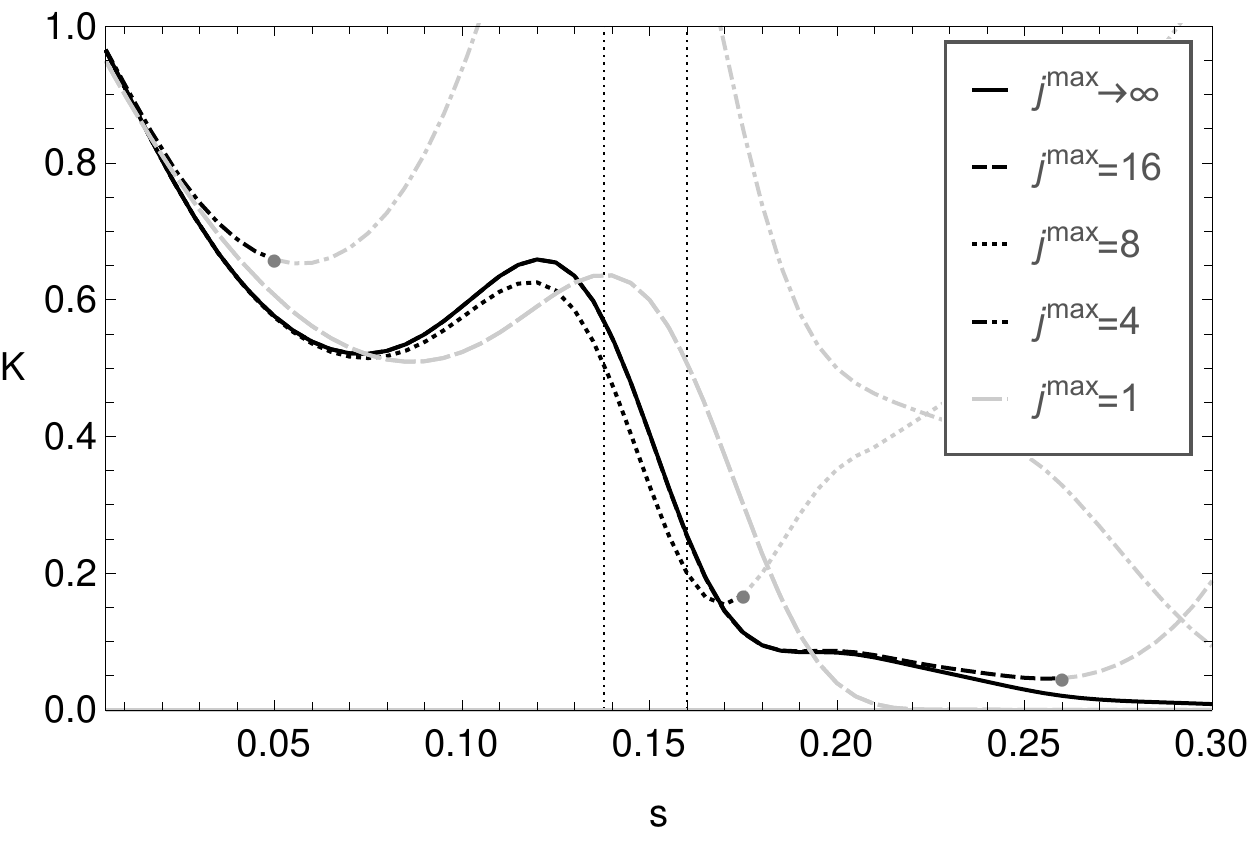}
 \caption{Perturbative expansion of the photon spectrum for a $4$-cycle, $\xi=0.5$ pulse colliding with an $\eta=0.1$ electron. $j^{\trm{max}}$ corresponds to the highest order included in $\xi$ from a truncated perturbation expansion of the amplitude. (Shading and plot points have been used to guide the eye to where the perturbation expansion breaks down.) $j^{\trm{max}}=1$ corresponds to linear Compton scattering. The higher the maximum perturbative order included, the higher the value of lightfront momentum, $s$, until which the expansion remains accurate. The vertical gridlines are the position of the nonlinear (left) and linear (right) Compton edge.} \label{fig:pertNew1}
\end{figure}

Before analysing the mid-IR peak, we first analyse the harmonic structure for the case $\xi=0.5$ (with $\eta=0.1$ and $N=16$), where we expect the perturbative expansion of $\tsf{K}$ to be accurate (since $\xi^2 < 1$). The results are presented in \figref{fig:pertNew1}. 

The first conclusion from \figref{fig:pertNew1} is that the harmonic order is not equivalent to how many background ``photons'' are interacting with the electron. In fact, even the $16$th order of charge-background coupling was not sufficient to reproduce the second harmonic. Rather, the harmonic order is the ``net'' number of photons \emph{absorbed} from the background. This is maybe unsurprising but worth emphasising. A harmonic expansion is often acquired using the Jacobi-Anger identity to rewrite exponentials of sinusoidal functions. For example, in a circularly-polarised monochromatic background, the terms linear in $a$ in \eqnref{eqn:Srprl1} contribute in the factor
\[
 \mbox{e}^{-i z \sin(\vphi-\psi)} = \sum_{n=-\infty}^{\infty} J_{n}(z)\mbox{e}^{-in(\vphi-\psi)},
\]
where $z=\xi |\mbf{r}^{\perp}|/\eta(1-s)$.

So we see that, expanding in $\xi$, even the first harmonic $j=1$ term is a sum over an infinite number of orders in $\xi$. Therefore, experiments that have measured second and third harmonics from nonlinear Thomson scattering (e.g. \cite{sakai15} where $\xi =0.5\ldots 0.7$ and \cite{khrennikov15} which observed a $\xi^2$ scaling of the spectral peak position for $\xi=0.83$) have already measured very high nonlinear orders of interaction between the electron and the laser background (as well as \cite{yan17}, which measured a nonlinear scaling in $\xi$ in the $\xi>1$ regime (for $\xi$ up to $\xi=12$), which included up to the $500\,\trm{th}$ harmonic).
\newline

\begin{figure}[h!!]
\centering
 \includegraphics[width=0.9\linewidth]{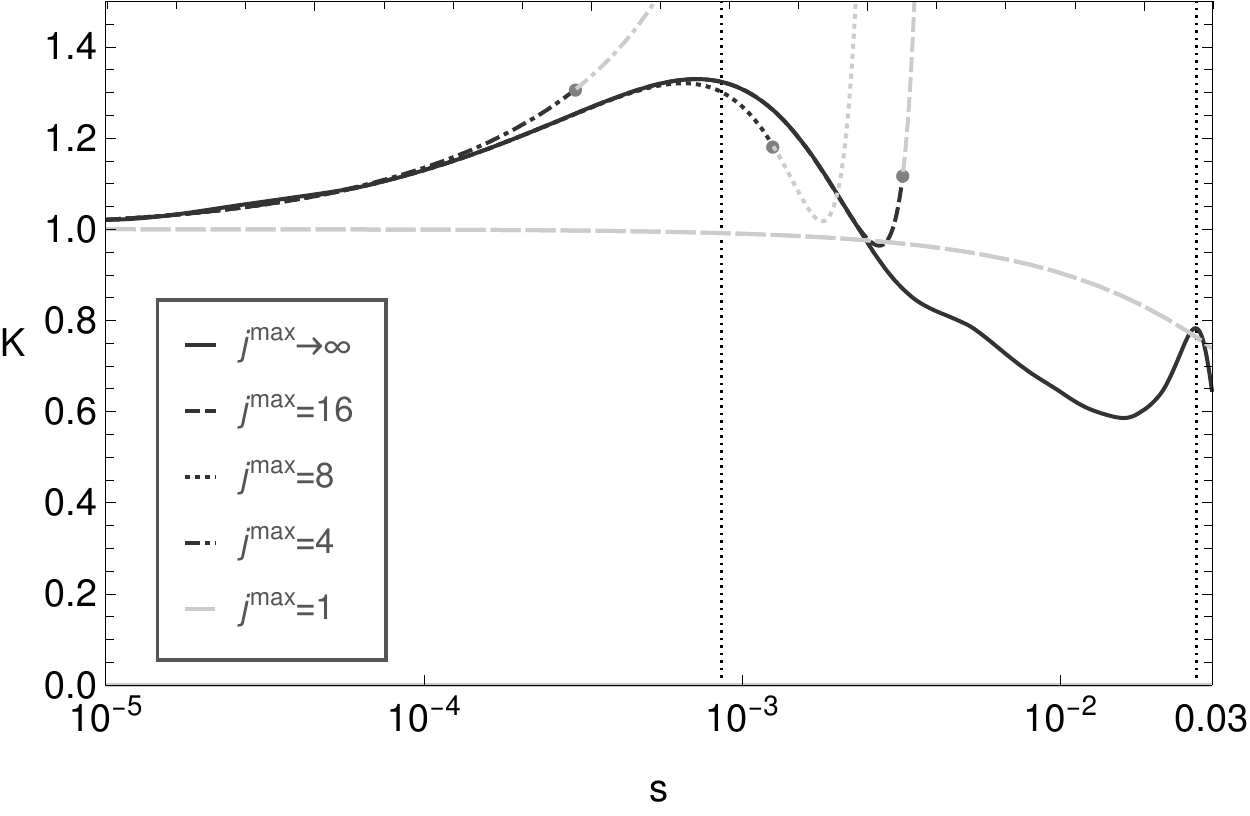}
 \caption{Perturbative expansion of the photon spectrum for a $\xi=2.5$, $16$-cycle pulse colliding with an $\eta=0.1$ electron, c.f. \figref{fig:pertNew1}. (Shading and plot points have been used to guide the eye to where the perturbation expansion breaks down.) The vertical gridlines are the estimate of the position of the mid-IR peak (left) and the first harmonic (right). 
 } \label{fig:pertNew2}
\end{figure}

Now we turn to the mid-IR peak and increase the value of the field strength to $\xi=2.5$. On the basis of \figref{fig:pertNew1}, one may na\"ively guess that, because $s$ is very small at the mid-IR peak ($s \approx \eta/N(1+\xi^2) \sim O(10^{-3})$), it can compensate for the fact that $\xi^2 > 1$, allowing a low-order perturbative expansion to suffice. However, in \figref{fig:pertNew2} we find the opposite. Including $16$ orders in the charge-field coupling is barely sufficient to reproduce the full mid-IR peak. As the mid-IR peak only appears when $\xi\gtrsim1$, we conclude that it is formed by a highly nonlinear interaction between the electron and laser pulse background, akin to the red-shifting of the Compton edge \cite{harvey09} and the generation of higher harmonics. This is the same way the mid-IR peak is formed in the flat-top model in \sxnref{sec:FT1}, where the $n=0$ harmonic becomes accessible due to the finite pulse duration, and corresponds to a photon being absorbed with  an energy of the order of the bandwidth of the laser pulse.

\section{Classical Analysis} \label{sec:CA}
Although highly nonlinear in the field interaction, the mid-IR peak must be a classical phenomenon. This is because the photon quantum parameter $\chi_\gamma = s \eta \xi$, which is a measure of the electron recoil when it emits a photon of momentum $s\eta$, obeys $\chi_\gamma \ll 1$. Therefore we can better understand the mid-IR peak, by asking: what part of the classical electron trajectory produces this spectral structure? The number of equivalent photons emitted classically is \cite{Peskin:1995ev}
\bea
\tsf{N}^{\trm{cl.}} = \int\frac{d^{3}k}{(2\pi)^{3}} \frac{1}{2k^{0}}|\tilde{\j}(k)|^{2}, \label{eqn:class1}
\eea
where the Fourier-transformed current in a plane wave background can be written \cite{jackson99}
\bea
 \tilde{\j}(k) = \frac{e}{\eta\,m^{2}}\int d\vphi~\Pi(\vphi)\,\mbox{e}^{\frac{i}{\eta} \int^{\vphi}_{-\infty} \frac{k\cdot \Pi(z)}{m^{2}} dz}. \label{eqn:class2}
\eea
Assuming $\tsf{N}^{\trm{cl.}}\ll 1$, the number of equivalent photons emitted classically can be interpreted as the classical limit for the probability of emission of a photon, $\tsf{P}^{\trm{cl.}}$. Now, the classical result will also \emph{diverge} when calculated na\"ively in a plane-wave pulse, and must be regularised, just as in QED. Regularisation of classical quantities is not a new concept (see e.g. \cite{dinu12}), but is relatively uncommon in the literature. We will see below how this regularisation is crucial to understanding the mid-IR peak. 

Regularisation of the classical probability results in the phase integral being performed only over the region where the plane wave background has support and with the momentum, $\Pi$, being replaced by $\Pi^{\trm{reg.}}$ where:
\[
\Pi^{\trm{reg.}} = \Pi - p \frac{k\cdot \Pi}{k\cdot p}.
\]
We note that whilst $k\cdot \Pi\neq0$, with the extra terms from regularisation: $k\cdot\Pi^{\trm{reg.}}=0$, which is required for current conservation ($k\cdot \tilde{\j}(k)=0$). Then the quantity that is relevant for the radiation spectrum is  $\Pi^{\trm{reg.}}(\vphi)\cdot \Pi^{\trm{reg.}}(\vphi')$, which we can write as:
\bea
&&\underbrace{\Delta(\vphi)\Delta(\vphi')}_{\trm{zero-field reg.}} + \underbrace{a(\vphi)\cdot a(\vphi')}_{\perp~\trm{term}} + \nn \\&& \underbrace{-\left[\frac{a^{2}(\vphi')}{2}\Delta(\vphi) +\frac{a^{2}(\vphi)}{2}\Delta(\vphi')\right]}_{\parallel~\trm{term}},
\label{eqn:decompF} 
\eea
where the symbols $\parallel$ ($\perp$) refer to momentum directions parallel (perpendicular) to the plane wave pulse propagation direction. 
\begin{figure}[h!!]
\centering
 \includegraphics[width=0.9\linewidth]{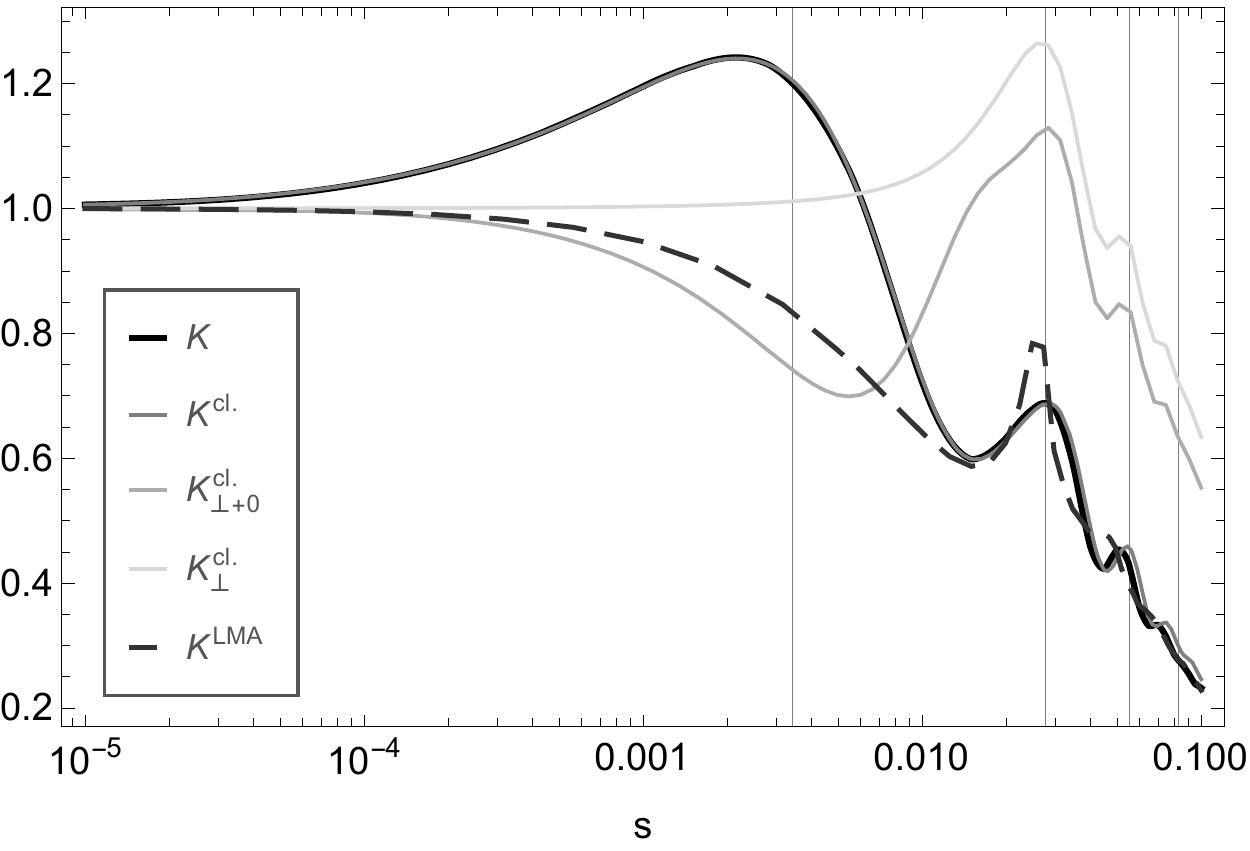}
 \caption{Comparison of different contributions to classical radiation spectrum ($\xi=2.5$, $\eta=0.1$, $N=4$) by removing different parts of the squared momentum in (\ref{eqn:decompF}). $K^{\trm{cl.}}_{\perp+0}$ corresponds to removing the longitudinal ``$\parallel$ term'' in (\ref{eqn:decompF}), whereas  $K^{\trm{cl.}}_{\perp}$ is for where the longitudinal and zero-field terms having been removed. The LMA, which misses the mid-IR peak, is plotted for comparison. The vertical gridlines are the harmonics and the approximation to the position of the mid-IR bump.} \label{fig:Kclass1}
\end{figure}
In \figref{fig:Kclass1} we illustrate the different spectra that are produced by artificially turning off different parts of the kinetic momentum. (The method of computation for the classical calculation is briefly detailed in \ref{sec:appCl}.) First of all, we confirm that, as expected, the mid-IR peak is well-approximated by a classical approach. Second, we find that the mid-IR peak disappears when we neglect the cross-term, which is the only place the longitudinal momentum gained by the electron from the background, $ (\xi^{2}(\vphi)/2\eta) \vkap$, survives. The acceleration of the electron due to this term is:
\bea
 \frac{d\Pi^{\parallel}_{\trm{reg.}}}{d\tau} = \frac{m}{2}\frac{d}{d\vphi}\xi_{\trm{env.}}^{2}(\vphi)\,\vkap, \label{eqn:acc1}
\eea
where with $\xi_{\trm{env.}}(\vphi)$, we emphasise that for the plane-wave pulse discussed here, the co-ordinate dependence of $\xi^{2}(\vphi)$ originates entirely from the pulse envelope ($\xi_{\trm{env.}}(\vphi) \equiv \xi(\vphi)$ for a circularly-polarised pulse). Let us contrast this with the monochromatic or even locally-monochromatic approach used where the ``instantaneous'' value of the pulse envelope is included, but higher-order derivatives are neglected \cite{Heinzl:2020ynb}. In both these cases, the acceleration in \eqnref{eqn:acc1} is zero, and there is no mid-IR peak. Therefore, we associate the mid-IR peak with this acceleration in the direction of propagation of the laser pulse, proportional to the gradient of the field squared. The mid-IR peak is then seen to originate from a \emph{pondermotive} effect over the scale of the pulse envelope, but in the longitudinal direction.

\section{Angular spectrum}
Since there is a separation in the $s$-spectrum between the mid-IR peak and the usual harmonic structure, a natural question is whether there is also a separation in the angular spectrum. Using \eqnref{eqn:pcon}, we can estimate the emission angle by taking the ratio of transverse to lightfront momentum of the scattered photon. We find:
\bea
 \frac{|\mbf{r}^{\perp}|^{2}}{s^{2}\eta^{2}} = \frac{2\bar{n}}{\eta}\frac{1-s}{s} - \frac{1}{\eta^{2}} \geq 0; \qquad \bar{n}\geq\frac{s}{2\eta(1-s)} \label{eqn:rp1}
\eea
where the second inequality follows from the first. \eqnref{eqn:rp1} tells us that:  i) as $s\to 0$, $\bar{n}$ is bounded below by $0$; ii) as $s\to0$, $|\mbf{r}^{\perp}|\to 0$. Therefore the mid-IR peak will feature at the centre of the angular distribution.

\begin{figure}[h!!]
\centering
 \includegraphics[width=0.85 \linewidth]{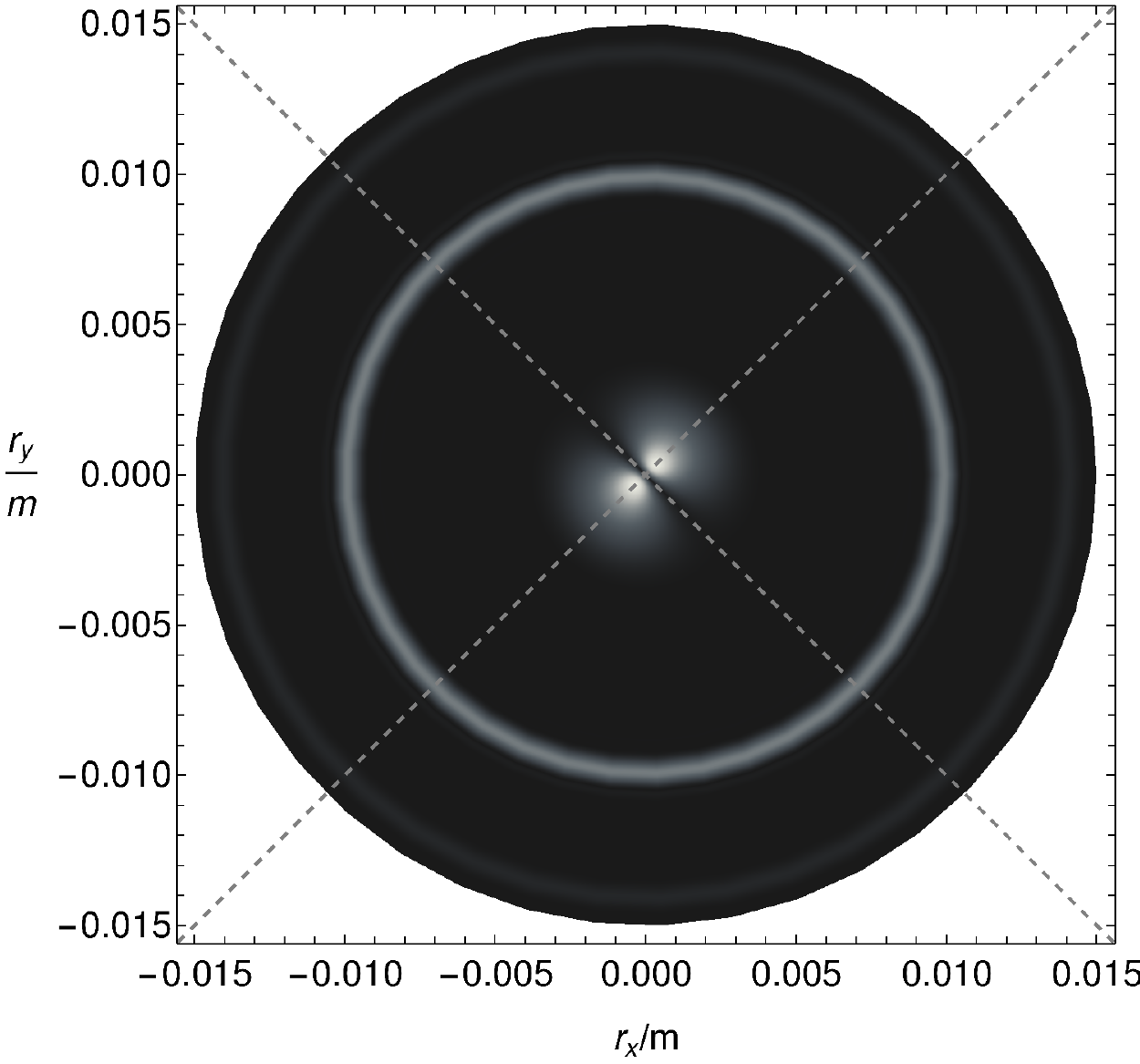}
 \caption{A plot of the square-root of the emission spectrum in the transverse, $\mbf{r}^{\perp}$, plane, for a $16$-cycle $\xi=2.5$ pulse colliding with an $\eta=0.1$ electron producing $s=5\times 10^{-4}$ photons (for a head-on collision with $1.55\,\trm{eV}$ laser photons, this corresponds to $4\,\trm{MeV}$ Compton-scattered photons). The outer ring is part of the first harmonic, whereas the inner lobes are the linearly polarised signal from the mid-IR peak. Dashed lines are plotted at angles $\pm\pi/4$ to the vertical.} \label{fig:angplotpol}
\end{figure}

Since $s \propto \eta_{\gamma} := s\eta = \vkap^{0}(k^0-k^\parallel)/m^2$, it is not immediately clear whether $k^\parallel$ is positive or negative (i.e. photons scattered parallel or antiparallel to the laser propagation direction). With some rearrangement, we arrive at:
\[
 k^\parallel = \frac{|\mbf{k}^{\perp}|^{2}-\kappa^2}{\kappa}; \quad \kappa := \frac{m^{2}\eta s}{\vkap^{0}}.
\]
We see the intuitive result that in the limit $\vkap^0/m \to 0$, $k^{\parallel}$ is negative, i.e. the photon is emitted antiparallel to the laser wavevector, and if $\vkap^0/m \to \infty$, the photon would turn out to be emitted parallel to the laser wavevector. To acquire $|\mbf{k}^{\perp}|^{2}$, we can calculate the angular spectrum of emitted photons. In \figref{fig:angplotpol}, we plot an example for $s=5\times10^{-4}$, $\xi=2.5$, $N=16$ and $\eta=0.1$.

We first note that the mid-IR peak can be clearly identified in the centre of the distribution. Suppose the collision between electron and laser is ``head-on'', then $\mbf{r}^{\perp} = \mbf{k}^{\perp}$. Here then $|\mbf{k}^{\perp}|/m\sim O(10^{-3})$, and $\eta s = 5 \times 10^{-5}$. If we take an optical laser pulse, e.g. at a wavelength of $800\,\trm{nm}$ (energy $1.55\,\trm{eV}$), then $\vkap^{0}/m \approx 3 \times 10^{-6}$, making $\kappa/m\sim O(10)$ i.e. $\kappa \gg |\mbf{k}^{\perp}|$ so that the photons in the mid-IR bump are scattered parallel to the electron and \emph{antiparallel} to the laser direction, very close to the propagation axis. (One could consider how high a laser frequency is required for the photons to be scattered \emph{parallel} to the laser direction in this case, and it would be around $25\,\trm{keV}$.)

To filter out the mid-IR signal from the photon spectrum, ideally, one would be able to apply both an angular and an energy cut to the photons detected. The angular cut would be used to select only those photons very close to the electron propagation axis, as in \figref{fig:angplotpol}, and the energy cut would be required otherwise at small values of $s$, the emission from the harmonics would overlap with emission due to the mid-IR peak. 

\begin{figure}[h!!]
\begin{flushleft}
 \fbox{{\scriptsize a) $\epsilon_{{\scriptsize +}}$-polarised}}\vspace{-0.25cm}
\end{flushleft}
\centering
 \includegraphics[width=0.485\linewidth]{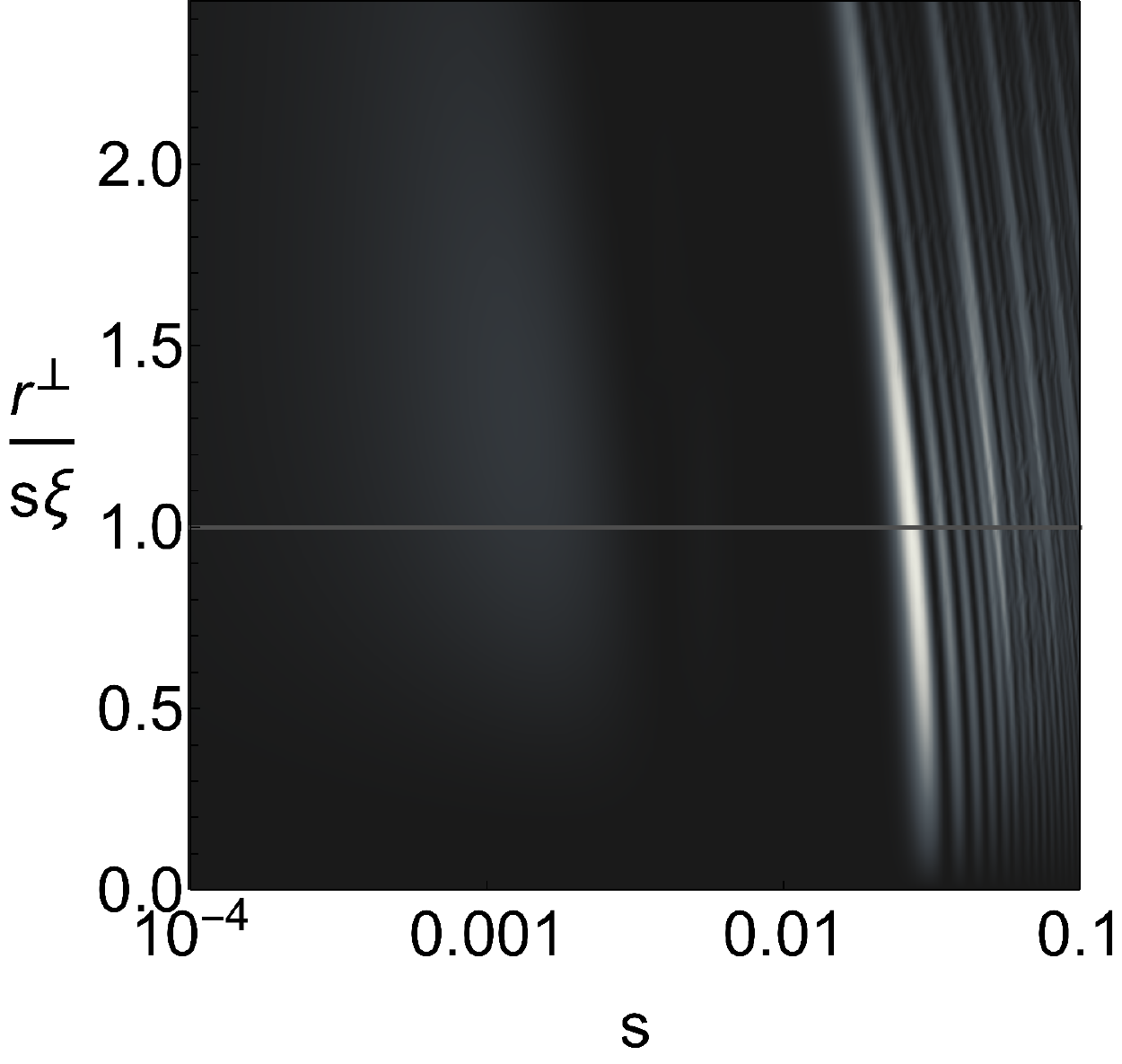}\includegraphics[width=0.485\linewidth]{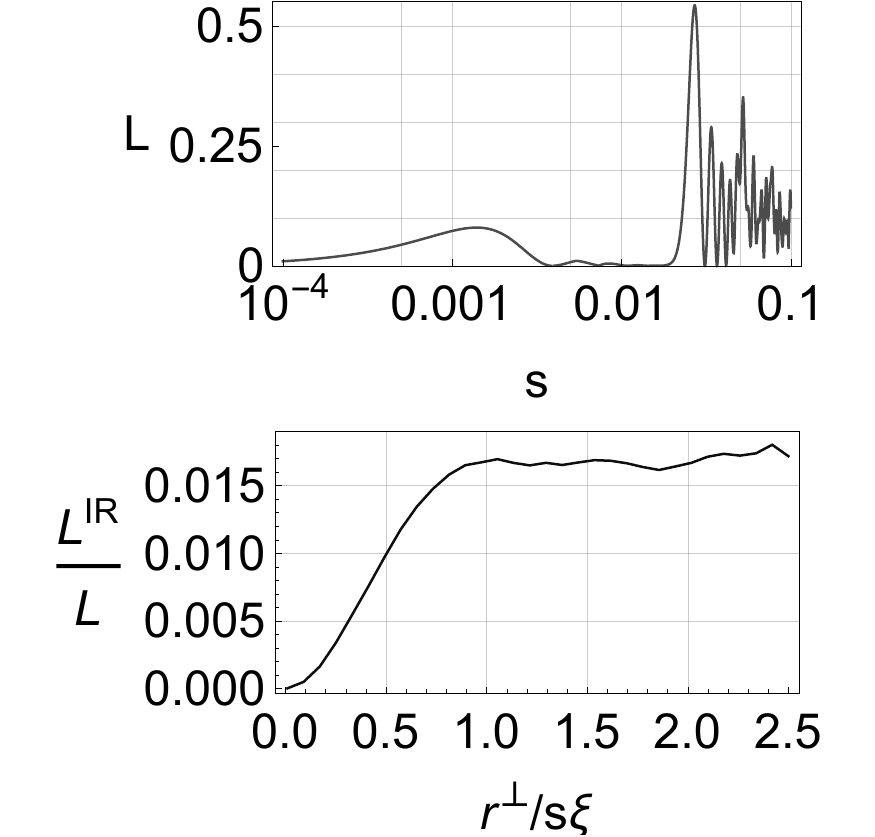}
\begin{flushleft}
 \fbox{{\scriptsize b) $\epsilon_{{\scriptsize -}}$-polarised}}\vspace{-0.25cm}
\end{flushleft}
\centering
 \includegraphics[width=0.485\linewidth]{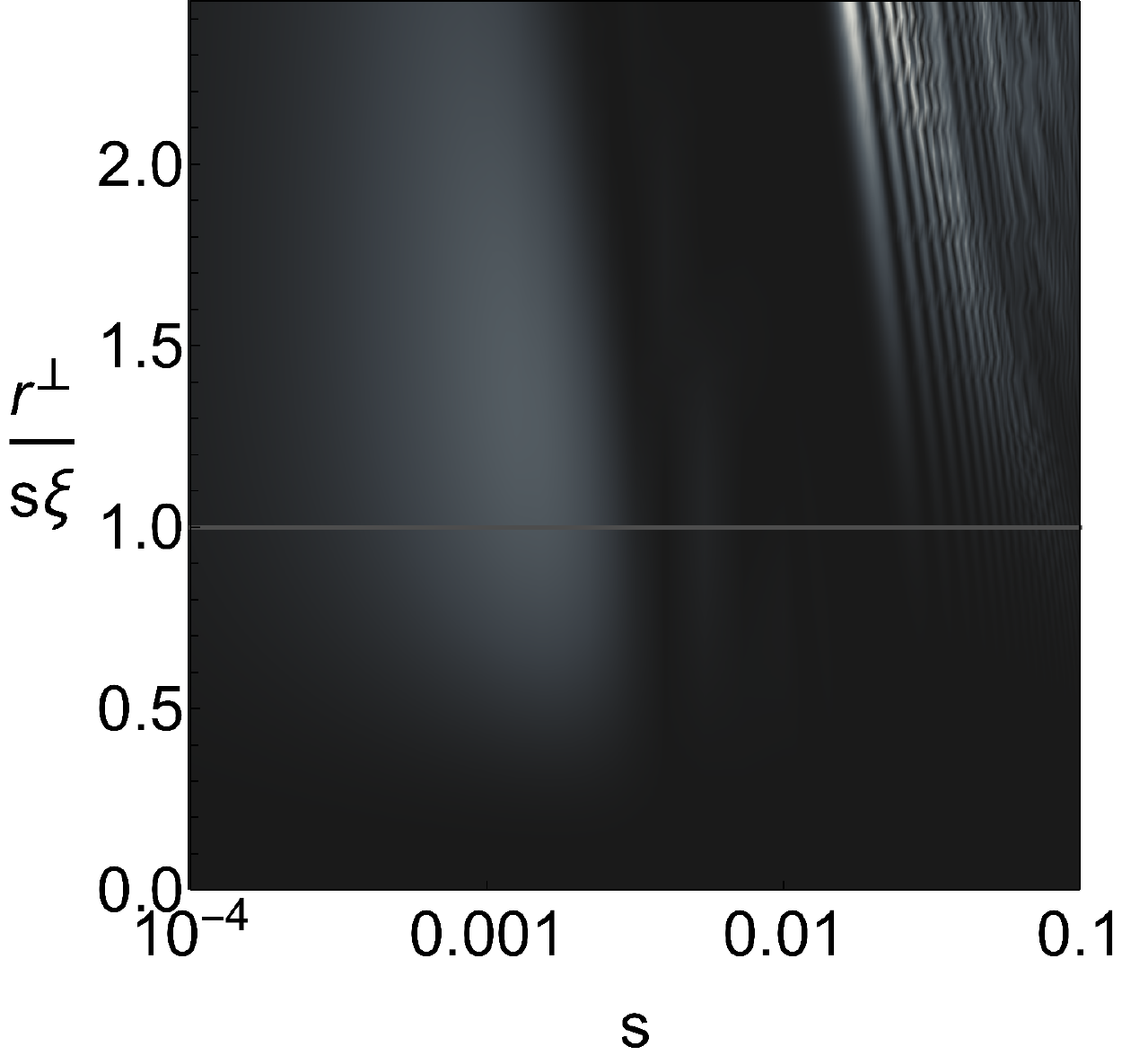}\includegraphics[width=0.485\linewidth]{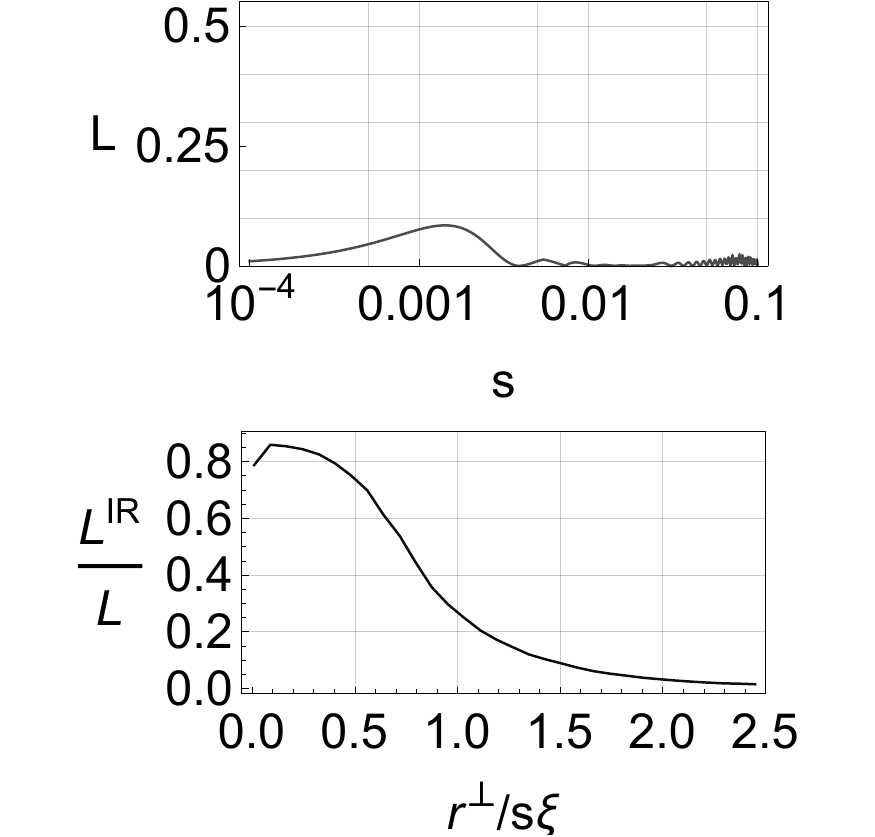}
 \caption{Angular spectrum for the collision of a $16$-cycle, $\xi=2.5$ pulse with an $\eta=0.1$ electron. Left: how the angular spectrum depends on photon polarisation (same scale). Right upper: a lineout of the angular spectrum, $L$, at $r^{\perp}=\xi$. Right lower: proportion of $s$-spectrum of photons originating from mid-IR peak, $L^{\trm{IR}}/L$, (defined as $s<0.01$) as a function of $r^{\perp}/\xi$.} \label{fig:angpol2}
\end{figure}

However, another difference between the mid-IR peak photons and the rest of the spectrum is apparent from \figref{fig:angplotpol}. In the centre of the distribution, a typical dipole distribution can be seen, showing that the mid-IR peak is clearly linearly polarised, whereas the harmonics are circularly-polarised. This occurs because the pulse envelope, which is responsible for the mid-IR peak, multiplies both transverse components of the background in the same way - there is no difference in the carrier-envelope phase between the two background polarisation directions. Therefore the envelope imposes a linearly-polarised structure on the background in contrast to the carrier frequency, which is circularly polarised. If one applies a filter to the scattered photons to separate right-handed from left-handed polarisation, then the harmonics can be partially filtered out. This is demonstrated in \figref{fig:angpol2} where the angular dependency is illustrated by calculating the quantity $\tsf{L}=d\tsf{K}/d(r^{\perp}/s)$.
We use the circular polarisation basis $\epsilon_{\pm} = (\epsilon_{1} \pm i\epsilon_{2})/\sqrt{2}$, where
\bea
\epsilon_{j} = \eps_{j}  - \vkap\, \frac{k\cdot \eps_{j}}{k\cdot \vkap}; \quad \eps^{\mu}_{j} = \delta_{j}^{\mu},~~ j\in\{1,2\}. \label{eqn:polbasis}
\eea
To calculate the spectrum of polarised photons, we do not sum over photon polarisation states, $l$, in \eqnref{eqn:I1}, but instead just select the polarisation required (recalling that $\epsilon^{\ast}$ enters the amplitude, and $\epsilon_{\pm}^{\ast} = \epsilon_{\mp}$).  From the results, we note that the mid-IR peak occurs with an equal amplitude in both $\epsilon_{\pm}$ circular polarisation states, which is logical because linearly-polarised photons are an equal mixture of circularly-polarised states. But we see that for $r^{\perp} \leq s\xi$, the signal from the harmonics is strongly suppressed in the $\epsilon_{-}$ polarised photons. This can be understood writing the background pulse, \eqnref{eqn:a1}, in terms of circular polarisation vectors:
\bea
a = \frac{m\xi}{\sqrt{2}}\,\sin^{2}\left(\frac{\vphi}{2N}\right)\left[\eps_{+}\mbox{e}^{-i\vphi} + \eps_{-}\mbox{e}^{i\vphi}\right]. \label{eqn:a2}
\eea
It will be sufficient to consider the linear Compton process. In this case, a photon must be \emph{absorbed} from the background, otherwise the process is kinematically forbidden - therefore only the $\eps_{+}$ polarisation is involved from the background. In the low-$s$ limit, the process should be well-approximated by the classical formula for Thomson scattering, the probability of which is $\propto (\epsilon_{\trm{in}}\cd \epsilon^{\ast}_{\trm{out}})^2$ \cite{jackson99}. For photon polarisation, we have chosen 
$\epsilon_\trm{out}= \epsilon_{\pm}$ from \eqnref{eqn:polbasis}. Suppose we consider photons being emitted mainly down the electron propagation axis, as we expect, because the electron is highly relativistic, and radiation is mainly emitted in a $\sim 1/\gamma$ emission ``cone'', then $\epsilon_{\pm} \approx \eps_{\pm}$. Then since $\epsilon_{\trm{in}} = \eps_{+}$, and since $\eps_{+}\cdot \eps_{-}^{\ast}=0$, we see that emission of $\eps_{-}$ polarised photons in the first harmonic, should be strongly suppressed, which is indeed what we find in \figref{fig:angpol2}. Therefore, in addition to an angular and energy cutoff, also a polarisation filter could be used to isolate the mid-IR signal photon \footnote{We mention for completeness sake that because the mid-IR peak corresponds to a small recoil parameter $s\chi\ll1$, the role of electron spin-flipping is negligible, which is also evidenced by how accurately the classical theory approximated the quantum theory.}.

\section{Conclusion}
In this work, the effect of the background's pulse envelope on the spectrum of nonlinear Compton scattering has been investigated. This corresponds to taking into account exactly, interference from emissions by the electron at different points in the pulse, up to the length scale of the entire pulse envelope. This was done, first in the toy model of a flat-top pulse, for which we presented the total probability as a sum over harmonics of an integral over two outgoing photon momentum components. The appearance of a spectral feature, which is absent in the well-known infinite or locally-monochromatic cases, of a harmonic peak in the mid infra-red (IR), was identified. This mid-IR peak was found to originate from the finite bandwidth of the flat-top pulse, which allowed the ``zeroth'' harmonic to become kinematically accessible. One can make an analogy, as was done for pair-creation in \cite{heinzl10}, of processes in a flat-top pulse being akin to emission from a diffraction grating. Here, the integer harmonics of the photon spectrum correspond to interference between the grating slits (provided by the carrier frequency of the pulse) whereas the zeroth and lower harmoniccs are due to interference from the finite width of the grating itself (provided by the flat-top pulse envelope). Opening of kinematic channels due to a laser pulse's finite bandwidth is known in other strong-field QED calculations such as pair-creation \cite{heinzl10} where, in \cite{Kohlfurst:2017hbd}, using the Dirac-Heisenberg-Wigner formalism a ponderomotive effect on the pair-spectrum was identified, real photon-photon scattering \cite{king12,Gies:2016czm,Gies:2017ezf,PhysRevA.98.023817} and the linear trident process \cite{Acosta:2019bvh}. This effect, which widens harmonic fringes in the emitted photon phase space is distinct from the opening of channels in a pulse due to just having a spacetime-dependent intensity and a variable effective mass. For this reason, the mid-IR peak is missed by local approximations, such as the locally constant field approximation (LCFA) and locally monochromatic approximation (LMA).

The rest of the paper analysed the mid-IR peak in the context of a more realistic plane-wave pulse with a sine-squared envelope. The mid-IR peak was found to be associated with a background wavevector approximately $1/2N$ smaller than from the carrier frequency, where $N$ is the number of cycles of the carrier frequency. The mid-IR peak only appears when the intensity parameter, $\xi$ (or ``$a_{0}$'') fulfills $\xi \gtrsim 1$, and its height in the lightfront spectrum grows as $\propto \xi^{3}$ compared to the IR limit, which is given by the leading-order term in $\xi$ and grows as $\propto \xi^{2}$. The mid-IR peak is a signature of an all-order interaction between the field and the charge, which was confirmed by the failure of a truncated perturbation expansion to approximate this part of the spectrum. Using a classical analysis, we were able to show that this peak arises from the ponderomotive force from the leading and trailing edges of the pulse envelope imparting a change in longitudinal momentum of the electron. Since this force is proportional to the \emph{derivative} of the envelope squared, this is an alternative explanation for why the mid-IR peak is missed from local approaches such as the LCFA and LMA, which neglect derivatives of the pulse envelope. Central to this analysis, was the regularisation of the zero-field limit of the \emph{classical} result. Finally, it was shown that if the carrier frequency is circularly-polarised, then the mid-IR peak, which comprises linearly-polarised photons, can be partially isolated from the rest of the spectrum using polarimetry and angular cuts of low-energy photons.

\subsection*{Acknowledgments}
BK thanks A. Ilderton for helpful discussions and comments on the manuscript. The author is supported by the EPSRC, Grant No.~EP/S010319/1.

\appendix
\section{Classical spectrum calculation} \label{sec:appCl}
Beginning with \eqnrefs{eqn:class1}{eqn:class2}, without loss of generality, we set $p\cdot\epsilon_{1,2}=0$. To obtain the spectrum, we then have:
\bea
 \tsf{K}^{\trm{cl.}} = \frac{8}{3\pi}\frac{\alpha}{N\xi^{2}\eta}\frac{1}{(2\pi)^{2}} \int d^{2}\mbf{r}^{\perp}\,\frac{T}{m^{2}},
\eea
where the pre-factor $8/3\pi N \xi^{2}$ is chosen so that \mbox{$\lim_{s\to0}\tsf{K}^{\trm{cl.}}=1$} and 
\bea
 T &=& \underbrace{|\tsf{S}^{\trm{cl.}}_{0,\Delta}|^{2}}_{\trm{zero-field reg.}} + \underbrace{\tsf{Re}\,\tsf{S}^{\trm{cl.}}_{2}\tsf{S}^{\trm{cl.}\,\ast}_{0,\Delta}}_{\parallel \trm{ term}} + \nn \\
 &&\underbrace{- |\tsf{S}^{\trm{cl.}}_{1\eps}|^{2} - |\tsf{S}^{\trm{cl.}}_{1\beta}|^{2}}_{\perp \trm{ term}} \label{eqn:classTdef}
\eea
where
\[
\tsf{S}^{\trm{cl.}}_{j} = \int_{\vphi_{i}}^{\vphi_{f}}\!\!d\vphi~ I_{j}(\vphi)\, \mbox{e}^{i(1-s)f},
\]
where $I_{0,\Delta}=\Delta$, $I_{1\epsilon}=-a\cdot\epsilon_{1}$,  $I_{1\beta}=-a\cdot\epsilon_{2}$ and ${I_{2}=-a\cdot a}$. The labelling of the terms in \eqnref{eqn:classTdef} corresponds to the labelling in the main text in \eqnref{eqn:decompF}. The exponent function $f$, is defined in the main text in \eqnref{eqn:Srprl1} and the regularisation factor $\Delta=1-k\cd\Pi/k\cd p$.
\clearpage

\bibliography{master}
\end{document}